\documentclass[a4paper,12pt]{article}

\usepackage{amsmath}
\usepackage{amssymb}
\usepackage{slashed}
\usepackage{amsthm}
\usepackage{xcolor}
\usepackage{calc}

\numberwithin{equation}{section}
\def\appendix#1{\addtocounter{section}{1}\setcounter{equation}{0}
\renewcommand{\thesection}{\Alph{section}}
\section*{Appendix \thesection\protect\indent \parbox[t]{11.15cm}{#1}}
\addcontentsline{toc}{section}{Appendix \thesection\ \ \ #1}}

\usepackage{accents}

\def\ciD{{\buildrel{\circ} \over \nabla}}

\newcommand{\bea}{\begin{eqnarray}}
\newcommand{\eea}{\end{eqnarray}}

\begin{document}

\begin{titlepage}
\begin{center}

\vspace*{-1.0cm}

\hfill  DMUS-MP-22-05
\\
\vspace{2.0cm}

\renewcommand{\thefootnote}{\fnsymbol{footnote}}
{\Large{\bf  $D=11$ $dS_5$ backgrounds with enhanced supersymmetry}}
\vskip1cm
\vskip 1.3cm
D. Farotti and J. Gutowski
\vskip 1cm
{\small{\it
Department of Mathematics,
University of Surrey \\
Guildford, GU2 7XH, UK.}\\
\texttt{d.farotti@surrey.ac.uk, j.gutowski@surrey.ac.uk}}

\end{center}
\bigskip
\begin{center}
{\bf Abstract}
\end{center}
We classify all warped $dS_5$ backgrounds in $D=11$ supergravity with enhanced supersymmetry. We show that backgrounds preserving $N=16$ supersymmetries consist of either a stack of M5 branes with transverse space $\mathbb{R}^5$,
or a generalized M5-brane configuration with transverse space $\mathbb{R} \times N_4$,
where $N_4$ is a hyper-K\"ahler manifold and the $M5$-brane harmonic function is determined by a hyper-K\"ahler potential on $N_4$. Moreover, we find that there are no backgrounds preserving exactly $N=24$ supersymmetries. Backgrounds preserving $N=32$ supersymmetries correspond to either $\mathbb{R}^{1,10}$ or $AdS_7\times S^4$.

\end{titlepage}

\section{Introduction}

Supersymmetry enhancement is known to play a particularly important role in the context of the geometric properties of supersymmetric black holes. It
has been shown that for many supergravity theories, the near-horizon limits
of supersymmetric extremal black holes{\footnote{In $N=2$, $D=4$ and $N=2$, $D=5$ supergravity, supersymmetric black holes are automatically extreme, however in
$D=11$ supergravity this need not be the case.}} (and also branes) exhibit supersymmetry enhancement, 
which in turn imposes additional conditions on the geometry near to the event horizon.
 In particular,  as a consequence of the enhanced supersymmetry, the isometry algebra enlarges in the near-horizon limit, containing a subalgebra isomorphic to $\mathfrak{sl}(2, \mathbb{R})$. In the context of $D=11$ and type IIA (including massive IIA) supergravity, it has been shown that near-horizon geometries with smooth fields preserve an even number of supersymmetries \cite{Gutowski:2013kma, Gran:2014fsa, Gran:2014yqa}.  The proof for this utilizes Lichnerowicz type theorems for certain generalized Dirac operators defined on the horizon spatial cross-section, which is assumed to be compact and without boundary. The index of these Dirac operators vanishes, which then establishes the supersymmetry enhancement.

Alternatively, one may consider the construction of a classification of highly supersymmetric solutions. In theories such as $D=11$ supergravity,
solutions preserving the minimal $N=1$ supersymmetry have rather weak
conditions on the geometry \cite{Gauntlett:2002fz, Gauntlett:2003wb}.
In contrast, it is reasonable to expect that the classification of
solutions with many supersymmetries will produce a much more restricted set
of geometries. An important result is the homogeneity theorem, which states that backgrounds preserving $N>16$ supersymmetry are locally homogeneous \cite{Figueroa-OFarrill:2012kws}, i.e. the tangent space at each point is spanned by the Killing vectors which are constructed as bilinears of the Killing spinors. The theorem has been proven for $D=11$ supergravity and type II $D=10$ supergravities, and holds for many other theories as well. Using an adaptation of the homogeneity theorem, combined with an analysis of the associated superalgebras, it has been shown that there are no $N>16$ smooth near-horizon geometries with non-trivial fluxes and also no warped $AdS_2$ backgrounds in ten or eleven dimensions \cite{Gran:2017qus}. This non-existence theorem applies
provided that the horizon section, or the internal manifold, respectively,
are compact and without boundary. Moreover, in \cite{Figueroa-OFarrill:2011tnp} homogeneous $D=11$ backgrounds which are symmetric have been classified up to local isometry. This provides a classification of $N>16$ symmetric backgrounds in $D=11$ supergravity. Furthermore, using spinorial geometry techniques it has been shown that all $D=11$ backgrounds preserving $N=30,31$ supersymmetries and all type IIB backgrounds preserving $N>28$ supersymmetries are maximally supersymmetric \cite{Gran:2006cn, Gran:2010tj,Gran:2007eu}; also there is a unique
plane wave solution in IIB supergravity preserving $N=28$ supersymmetry \cite{Gran:2009cz}. The spinorial geometry technique is a powerful tool to solve the Killing spinor equations (KSE) of supergravity theories and can be adapted to backgrounds with near maximal number of supersymmetries. It is based on the use of the gauge covariance of the KSE, together with a representation of the Clifford algebra, in an appropriate oscillator basis, acting on spinors which correspond to differential forms \cite{Gillard:2004xq, Gran:2005wu}.

In this paper, we shall classify the warped product $dS_5 \times_w M_6$ solutions in $D=11$ supergravity which exhibit enhanced supersymmetry. In \cite{Farotti:2022xsd} it was
shown that supersymmetric warped product $dS_5 \times_w M_6$ solutions must preserve 
$N=8k$ supersymmetries for $k=1,2,3,4$. Minimal supersymmetry therefore corresponds to
$N=8$ supersymmetry - these were fully classified in \cite{Farotti:2022xsd}. 
Furthermore, it was also noted that the only possible $N=32$ warped-product
$dS_5 \times_w M_6$ solutions are $\mathbb{R}^{1,10}$ with vanishing 4-form, or the maximally supersymmetric $AdS_7 \times S^4$ solution. Hence, in this paper we shall primarily be
concerned with the $N=16$ and $N=24$ cases. We shall prove that there are no
exactly $N=24$ solutions. This is somewhat analogous to the analysis for warped
product $AdS_5 \times_w M_6$ solutions in \cite{Beck:2016lwk}, in which it was proven
that there are no $N=24$ warped product $AdS_5$ solutions. 
However, there are differences between the $AdS_5$ and $dS_5$ analysis. The non-existence of
$N=24$ $AdS_5$ solutions was established via 
an adapted version of the homogeneity theorem in \cite{Figueroa-OFarrill:2012kws}, together
with a maximum principle argument which utilizes certain (assumed) global properties of
the internal space.
In contrast, for the $dS_5$ solutions, the local geometric conditions are sufficiently strong to allow an
explicit integration of the Killing spinor equation along two of
the directions of $M_6$. The analysis of the Killing spinor equation then
simplifies significantly to the counting of certain parallel spinors
on a hyper-K\"ahler manifold. This enables the case of $N=24$ supersymmetries to
be excluded by direct inspection. Moreover, this notable simplification also
allows for the full classification of the $N=16$ $dS_5$ solutions. 
We find two classes of $N=16$ $dS_5 \times_w M_6$ backgrounds. The first is a special class of solutions constructed
in \cite{Farotti:2022xsd} for which the 4-form is parallel; the geometry
is a generalized M5-brane configuration with transverse space $\mathbb{R} \times N_4$,
where $N_4$ is a hyper-K\"ahler manifold and the $M5$-brane harmonic function is determined by a hyper-K\"ahler potential on $N_4$. The second class is a stack of M5 branes with transverse space $\mathbb{R}^5$ \cite{Gueven:1992hh}. Further recent progress towards classifying the $N=16$ $AdS_5$ warped product solutions 
has been made in \cite{Papadopoulos:2020mbw}, which develops a systematic examination of spacetimes, and other fields, which are invariant under the action of certain R-symmetry groups.

It is known that there are many no-go theorems which imply non-existence
of de Sitter solutions in supergravity \cite{gbds, deWit:1986mwo, Maldacena:2000mw},
in cases for which the warp factor and fluxes and are smooth, and the internal manifold is smooth and compact without boundary.
Our motivation is therefore to construct a systematic classification of supersymmetric
de Sitter solutions in supergravity theories from a purely local perspective. In addition,
we shall not assume that the spinors factorize into products of spinors on $dS_n$
and on the internal space, as it is known that such factorizations can produce a miscounting of supersymmetries \cite{Gran:2016zxk}. This classification programme
has already been fully completed in the case of heterotic supergravity, including the first order
corrections in $\alpha'$ \cite{Farotti:2022twf}. For heterotic warped product
de Sitter geometries, the warped product $dS_2$ solutions are in 1-1 correspondence with
the direct product $AdS_3$ solutions classified in \cite{Beck:2015gqa}; moreover all
warped product $dS_n$ solutions for $3 \leq n \leq 9$ are direct product $\mathbb{R}^{1,n} \times
M_{9-n}$ backgrounds. This is consistent with the restrictions on heterotic $dS_n$ solutions
for $n \geq 4$ found in \cite{Kutasov:2015eba}; it is also clear from 
the analysis of \cite{Farotti:2022twf} that the warped product $dS_2$
and $dS_3$ solutions are also highly restricted in heterotic supergravity.
The warped product $dS_n$ solutions in $D=11$ supergravity exhibit similar foliation properties for
$5 \leq n \leq 10$. In the case of $n=5$, $dS_5$ arises as a (conformal) foliation of $\mathbb{R}^{1,5}$, corresponding to the directions along the M5-brane worldvolume.
 In contrast, the warped product $dS_4$ solutions with minimal $N=8$ supersymmetry have been classified in \cite{DiGioia:2022bqg}, and there is no analogous foliation of $dS_4$ into $AdS_5$ or $\mathbb{R}^{1,4}$ appearing.

%

The plan of this paper is as follows: in section 2 we summarize some key aspects of the
classification of warped product $dS_5$ solutions, $dS_5 \times_w M_6$, preserving the minimal amount of $N=8$ supersymmetry constructed in \cite{Farotti:2022xsd}, in which it is shown that all
such solutions are generalized $M5$-brane solutions for which the 5-dimensional transverse 
manifold is $\mathbb{R} \times N_4$, where $N_4$ is a hyper-K\"ahler manifold. 
A particularly useful special case, which turns out to have enhanced supersymmetry, for which the 4-form $F$ is covariantly constant,
is also presented. In section 3, we use the results presented in section 2 to 
explicitly integrate the Killing spinor equations, acting on a generic spinor, along $M_6$.
This produces a gravitino type equation, and an algebraic condition. The analysis then
splits into two subcases, depending on the properties of the algebraic condition. In each
of these subcases, the Killing spinor equations can ultimately be reduced to counting
parallel spinors on the hyper-K\"ahler manifold $N_4$, enabling
all of the solutions with enhanced supersymmetry to be fully classified. The results
are summarized in section 4.

\section{$D=11$ warped product $dS_5$ backgrounds}

In this section, we summarize the key results about warped $dS_5\times_w M_6$ backgrounds in $D=11$ supergravity \cite{Farotti:2022xsd}, which were derived for solutions preserving the minimal $N=8$ supersymmetry. First of all, the 11-dimensional metric is given by
\begin{eqnarray}
ds^2(M_{11})=A^2ds^2(dS_5)+ds^2(M_6)
\label{11dmetric}
\end{eqnarray}
where $A$ is a function of the co-ordinates of the internal Riemannian manifold $M_6$ and
\begin{eqnarray}
ds^2(dS_5)=\frac{1}{\big(1+\frac{k}{4}|x|^2\big)^2}\eta_{\mu\nu}dx^{\mu}dx^{\nu}~,~~~\mu,\nu=0,1,\dots, 4
\end{eqnarray}
is the metric of 5-dimensional de-Sitter spacetime, with $|x|^2=\eta_{\mu\nu}x^{\mu}x^{\nu}$ and $k=\frac{1}{\ell ^2}$. We require that the Lie derivative of the 4-form flux $F$ with respect to all of the isometries of $dS_5$ vanishes, and consequently
\begin{eqnarray}
F = X
\end{eqnarray}
where $X$ is a 4-form on $M_6$ whose components depend only on the co-ordinates of 
$M_6$. Moreover, as we have mentioned previously, in what follows we do not make any assumption on smoothness of $A$ or the 4-form flux , neither do we require the internal manifold to be smooth or compact without boundary. Rather, the analysis is entirely local. \\
\indent
Let us introduce the space-time vielbein
\begin{eqnarray}
\textbf{e}^{\mu}=\frac{A}{1+\frac{k}{4}||x||^2}dx^{\mu}~,~~~~~\textbf{e}^a=e^a_{~\alpha}(y)dy^{\alpha}
\label{viel1}
\end{eqnarray}
where $a=5,6,\dots ,\sharp$ is a frame index on $M_6$ and we have denoted by $y^{\alpha}$ the co-ordinates on $M_6$. The bosonic field equations of $D=11$ supergravity
\begin{eqnarray}
R_{AB}=\frac{1}{12}F_{AL_1L_2L_3}F_B^{~L_1L_2L_3}-\frac{1}{144}g_{AB}F^2
\end{eqnarray}
\begin{eqnarray}
d\star_{11}F-\frac{1}{2}F\wedge F=0
\label{4formEOMS}
\end{eqnarray}
and the Bianchi identities
\begin{eqnarray}
dF=0
\end{eqnarray}
can be reduced on $M_6$ yielding
\begin{eqnarray}
4kA^{-2}-A^{-1}\widetilde{\nabla}^a\widetilde{\nabla}_a A-4A^{-2}(\widetilde{\nabla}A)^2+\frac{1}{12}G^2=0
\label{eins456}
\end{eqnarray}
\begin{eqnarray}
\widetilde{R}_{ab}=5A^{-1}\widetilde{\nabla}_a\widetilde{\nabla}_b A+\frac{1}{6}G^2\delta_{ab}-\frac{1}{2}G_{cb}G^c_{~a}
\label{eins457}
\end{eqnarray}
\begin{eqnarray}
\tilde{d}(A^5 G)=0
\label{gaugeds5}
\end{eqnarray}
\begin{eqnarray}
\tilde{d}\star_6 G=0
\label{bianchids5}
\end{eqnarray}
where $\widetilde{\nabla}$ denotes the Levi-Civita connection on $M_6$, $\tilde{d}$ is the exterior derivative on $M_6$, and
\begin{eqnarray}
G=\star_6 X~.
\label{gdef}
\end{eqnarray}
Furthermore, the KSE  of $D=11$ supergravity
\begin{eqnarray}
\bigg(\nabla_A-\frac{1}{288}\Gamma_A^{~~B_1B_2B_3B_4}F_{B_1B_2B_3B_4}+\frac{1}{36}F_{AB_1B_2B_3}\Gamma^{B_1B_2B_3}\bigg)\epsilon=0
\label{KSE1}
\end{eqnarray}
can be integrated along the de-Sitter directions yielding
\begin{eqnarray}
\epsilon=\big(1+\frac{k}{4}||x||^2\big)^{-\frac{1}{2}}\bigg(1+x^{\mu}\Gamma_{\mu}\big(-\frac{1}{2}\widetilde{\slashed{\nabla}}A+\frac{A}{288}\slashed{X}\big)\bigg)\psi
\end{eqnarray}
where $\psi$ is a 32-component Majorana spinor on $M_6$ which satisfies 
\begin{eqnarray}
\widetilde{\nabla}_a\psi=\bigg(-\frac{1}{12}G_{ab}\Gamma^b+\frac{1}{12}\Gamma_a^{~~bc}G_{bc}\bigg)\Gamma^{(7)}\psi
\label{killingG}
\end{eqnarray}
and
\begin{eqnarray}
\Gamma^{(7)}=\frac{1}{6!}\epsilon_{a_1a_2\dots a_6}\Gamma^{a_1a_2\dots a_6}
\end{eqnarray}
is the highest rank Gamma matrix on $M_6$. As shown in \cite{Farotti:2022xsd}, we can pick co-ordinates $\{t,s,z^i\}$, with $i=1,2,3,4$ on $M_6$ such that
\begin{eqnarray}
ds^2(M_6)=f^{-4}(s,z)ds^2+\frac{1}{k}f^2(s,z)dt^2+f^{-4}(s,z)ds^2(N_4)
\label{M6rescaleds}
\end{eqnarray}
where $N_4$ is a 4-dimensional hyperK\"ahler manifold, with metric tensor
\begin{eqnarray}
ds^2(N_4)=h_{ij}(z)dz^idz^j~.
\end{eqnarray}
Moreover, the warp factor $A$ is given by
\begin{eqnarray}
A=t\cdot f(s,z)
\label{Afine}
\end{eqnarray}
and the 2-form flux $G$ is
\begin{eqnarray}
G=\frac{6}{\sqrt{k}}df\wedge dt~.
\label{Gcoordinate}
\end{eqnarray}
Setting $f=H^{-{1 \over 6}}$, and using \eqref{M6rescaleds} and \eqref{Afine}, the 11-dimensional metric tensor \eqref{11dmetric} is given by
\begin{eqnarray}
ds^2(M_{11}) = H^{-{1 \over 3}} ds^2({\mathbb{R}}^{1,5}) + H^{{2 \over 3}} ds^2 ({\mathbb{R}} \times N_4)
\label{metricHH}
\end{eqnarray}
and the Einstein equation \eqref{eins456} simplifies to
\begin{eqnarray}
\Box_5 H =0
\label{harmonicH}
\end{eqnarray}
where $\Box_5$ denotes the Laplacian on ${\mathbb{R}} \times N_4$. Moreover, \eqref{Gcoordinate} yields
\begin{eqnarray}
F = \star_5 dH
\label{4formH}
\end{eqnarray}
where $\star_5$ is the Hodge dual on ${\mathbb{R}} \times N_4$. The geometry given by \eqref{metricHH}-\eqref{4formH} corresponds to that of a generalized M5-brane configuration, with transverse space ${\mathbb{R}} \times N_4$ \cite{Gauntlett:1997pk}.\\

\subsection{Solutions with parallel 4-form}

A sub-class of these solutions corresponds to those backgrounds for which the 4-form $F$ is covariantly constant with respect to the 11-dimensional Levi-Civita connection; as we shall prove in the section 3.1, this class of solutions actually has enhanced supersymmetry.
These backgrounds satisfy
\begin{eqnarray}
\widetilde{\nabla} G=0~.
\label{covconstG}
\end{eqnarray}
Note that \eqref{covconstG} implies
\begin{eqnarray}
G^2=c^2
\label{GC}
\end{eqnarray}
with $c$ constant. The condition \eqref{covconstG} yields the following set of PDEs
\begin{eqnarray}
\frac{\partial^2f}{\partial s^2}+f^{-1}\bigg(\frac{\partial f}{\partial s}\bigg)^2-2f^{-1}h^{ij}\frac{\partial f}{\partial z^i}\frac{\partial f}{\partial z^j}=0
\label{pde1}
\end{eqnarray}
\begin{eqnarray}
\frac{\partial f}{\partial s\partial z^i}+3f^{-1}\frac{\partial f}{\partial s}\frac{\partial f}{\partial z^i}=0
\label{pde2}
\end{eqnarray}
\begin{eqnarray}
\ciD_i \ciD_j f+3f^{-1}\frac{\partial f}{\partial z^i}\frac{\partial f}{\partial z^j}-2f^{-1}h_{ij}\bigg(\bigg(\frac{\partial f}{\partial s}\bigg)^2+h^{kl}\frac{\partial f}{\partial z^k}\frac{\partial f}{\partial z^l}\bigg)=0~.
\label{pde3}
\end{eqnarray}
Equations \eqref{pde1}-\eqref{pde3}, supplemented by \eqref{GC}, are equivalent to
\begin{eqnarray}
f(s,z)=\sqrt{2}\bigg(c^2s^2+P(z)\bigg)^{\frac{1}{4}}
\label{fszizi2}
\end{eqnarray}
\begin{eqnarray}
(\ciD P)^2=4c^2P
\label{usefulPP2}
\end{eqnarray}
\begin{eqnarray}
\ciD_i \ciD_j P=2c^2 h_{ij}
\label{usefulhij2}
\end{eqnarray}
where $\ciD$ denotes the Levi-Civita connection on $N_4$. In particular, by taking $c=0$ and $N_4=\mathbb{R}^4$, we recover the maximally supersymmetric solution $\mathbb{R}^{1,10}$ with vanishing 4-form.\footnote{The other maximally supersymmetric solution which is a warped product $dS_5$ solution is $AdS_7\times S^4$, which arises as the near-horizon limit of the standard M5-brane, with transverse space $\mathbb{R}^5$.} In particular, ({\ref{usefulhij2}}) implies that
${P \over 2c^2}$ is a hyper-K\"ahler potential for $N_4$, \cite{Swann}.

\section{Integration of the KSE on $M_6$}

In this section, we integrate the KSE \eqref{killingG} along two of the directions
of $M_6$, corresponding to the co-ordinates $s$ and $t$. The resulting reduced Killing spinor equations ultimately are, after some further manipulation, equivalent to 
requiring the existence of certain parallel spinors on $N_4$. As we shall see, this enables to classify $dS_5$ backgrounds with extended supersymmetry. First of all, let us introduce the vielbein on $M_6$
\begin{eqnarray}
\tilde{\textbf{e}}^5=f^{-2}ds ~,~~~\tilde{\textbf{e}}^6=\frac{1}{\sqrt{k}}f dt~,~~~\tilde{\textbf{e}}^I=f^{-2}\buildrel{\circ} \over e^I
\label{vielbeinm6}
\end{eqnarray}
where $\buildrel{\circ} \over e^I=\buildrel{\circ} \over e^I_{~i}dz^i$ and
\begin{eqnarray}
h_{ij}=\delta_{IJ}\buildrel{\circ} \over e^I_{~i}\buildrel{\circ} \over e^J_{~j}~.
\end{eqnarray}
Using \eqref{vielbeinm6}, equation \eqref{M6rescaleds} reads
\begin{eqnarray}
ds^2(M_6)=(\tilde{\textbf{e}}^5)^2+(\tilde{\textbf{e}}^6)^2+\delta_{IJ}\tilde{\textbf{e}}^I\tilde{\textbf{e}}^J~.
\end{eqnarray}
The non-vanishing components of the spin connection on $M_6$ with respect to the frame \eqref{vielbeinm6} are given by
\begin{eqnarray}
\widetilde{\Omega}_{6,65}=f\frac{\partial f}{\partial s}
\label{spinm61}
\end{eqnarray}
\begin{eqnarray}
\widetilde{\Omega}_{6,6I}=-\frac{1}{2}\widetilde{\Omega}_{5,5I}=f\buildrel{\circ} \over e_If
\end{eqnarray}
\begin{eqnarray}
\widetilde{\Omega}_{I,5J}=2f\frac{\partial f}{\partial s}\delta_{IJ}
\end{eqnarray}
\begin{eqnarray}
\widetilde{\Omega}_{I,JK}=f^2\buildrel{\circ} \over\Omega_{I,JK}-4f\delta_{I[J}\buildrel{\circ} \over e_{K]} f
\label{spinm6fin}
\end{eqnarray}
where $\buildrel{\circ} \over\Omega_{I,JK}$ is the spin connection on $N_4$.\\
\indent  
Using \eqref{spinm61}-\eqref{spinm6fin} and \eqref{Gcoordinate}, the KSE \eqref{killingG} read
\begin{eqnarray}
\frac{\partial\psi}{\partial t}=\frac{1}{2\sqrt{k}}\bigg(f^2\frac{\partial f}{\partial s}\Gamma^5+f^2\buildrel{\circ} \over \nabla_{I} f \Gamma^I\bigg)\big(\Gamma^6+\Gamma^{(7)}\big)\psi
\label{partialt}
\end{eqnarray}
\begin{eqnarray}
\frac{\partial \psi}{\partial s}=f^{-1}\buildrel{\circ} \over \nabla_{I} f\Gamma^6\Gamma^5\Gamma^I\big(\Gamma^{6}+\Gamma^{(7)}\big)\psi-\frac{1}{2}f^{-1}\frac{\partial f}{\partial s} \Gamma^6 \Gamma^{(7)}\psi
\label{partials}
\end{eqnarray}
\begin{eqnarray}
\buildrel{\circ} \over \nabla_I \psi&=&f^{-1}\frac{\partial f}{\partial s}\Gamma^6\Gamma_I \Gamma^5\big(\Gamma^6+\Gamma^{(7)}\big)\psi+f^{-1}\buildrel{\circ} \over \nabla_{J}f\Gamma^6\Gamma_I^{~~J}\big(\Gamma^6+\Gamma^{(7)}\big)\psi
\nonumber \\
&-&\frac{1}{2}f^{-1}\buildrel{\circ} \over \nabla_{I} f\Gamma^6\Gamma^{(7)}\psi~.
\label{partiali}
\end{eqnarray}
Since $f$ does not depend on $t$, equation \eqref{partialt} can be easily integrated, yielding
\begin{eqnarray}
\psi=e^{t\mathcal{X}}\eta~,~~~~~\frac{\partial\eta}{\partial t}=0
\label{psitx}
\end{eqnarray}
where $\eta$ is a 32-component Majorana spinor and
\begin{eqnarray}
\mathcal{X}:=\frac{1}{2\sqrt{k}}\bigg(f^2\frac{\partial f}{\partial s}\Gamma^5+f^2\buildrel{\circ} \over \nabla_{I} f \Gamma^I\bigg)\big(\Gamma^6+\Gamma^{(7)}\big)~.
\end{eqnarray}
Notice that $\mathcal{X}^2=0$, thus $e^{t\mathcal{X}}=1+t\mathcal{X}$ and \eqref{psitx} yields
\begin{eqnarray}
\psi=\eta+\frac{t}{2\sqrt{k}}\bigg(f^2\frac{\partial f}{\partial s}\Gamma^5+f^2\buildrel{\circ} \over \nabla_{I} f \Gamma^I\bigg)\big(\Gamma^6+\Gamma^{(7)}\big)\eta~.
\label{psitintegrated}
\end{eqnarray}
We can rewrite \eqref{partials}, \eqref{partiali} and \eqref{psitintegrated} covariantly on $\mathbb{R}\times N_4$ as follows
\begin{eqnarray}
D_{\alpha}\psi=f^{-1}D_{\beta}f\Gamma^6\Gamma_{\alpha}^{~~\beta}(\Gamma^6+\Gamma^{(7)})\psi-\frac{1}{2}f^{-1}D_{\alpha}f\Gamma^6\Gamma^{(7)}\psi
\label{KillingN51}
\end{eqnarray}
and
\begin{eqnarray}
\psi=\eta+\frac{t}{2\sqrt{k}}f^2D_{\alpha}f\Gamma^{\alpha}\big(\Gamma^6+\Gamma^{(7)}\big)\eta
\label{KillingN52}
\end{eqnarray}
where $D$ is the Levi-Civita connection on $\mathbb{R}\times N_4$ and $\alpha,\beta$ are frame indices on $\mathbb{R}\times N_4$. Inserting \eqref{KillingN52} into \eqref{KillingN51}, the vanishing of the $t$-independent terms and the terms linear in $t$ yields
\begin{eqnarray}
D_{\alpha}\eta=f^{-1}D_{\beta}f\Gamma^6\Gamma_{\alpha}^{~~\beta}(\Gamma^6+\Gamma^{(7)})\eta-\frac{1}{2}f^{-1}D_{\alpha}f\Gamma^6\Gamma^{(7)}\eta
\label{KillingN53}
\end{eqnarray}
and
\begin{eqnarray}
\mathcal{H}_{\alpha\beta}\Gamma^{\beta}(\Gamma^6+\Gamma^{(7)})\eta=0
\label{KillingN54}
\end{eqnarray}
respectively, where we have defined
\begin{eqnarray}
\mathcal{H}_{\alpha\beta}:=f^2D_{\alpha}D_{\beta}f+3fD_{\alpha}fD_{\beta}f-2f(Df)^2\delta_{\alpha\beta}~.
\label{definHIJ}
\end{eqnarray}
Two distinct cases must be considered, depending on whether $\mathcal{H}_{\alpha\beta}$ vanishes or not.

\subsection{$\mathcal{H}_{\alpha\beta}=0$}

If $\mathcal{H}_{\alpha\beta}=0$, then \eqref{KillingN54} is automatically satisfied and \eqref{definHIJ} implies
\begin{eqnarray}
f^2D_{\alpha}D_{\beta}f+3fD_{\alpha}fD_{\beta}f-2f(Df)^2\delta_{\alpha\beta}=0~.
\label{KillingN55}
\end{eqnarray}
Decomposing \eqref{KillingN55} on $N_4$, we find \eqref{pde1}-\eqref{pde3}, hence $\widetilde{\nabla} G=0$ and equations \eqref{fszizi2}-\eqref{usefulhij2} hold.
In the following, it is convenient to write $\eta$ as
\begin{eqnarray}
\eta=\eta^+ +\eta^-
\label{chir20}
\end{eqnarray}
where 
\begin{eqnarray}
\Gamma^6\Gamma^{(7)}\eta^{\pm}=\pm \eta^{\pm}~.
\label{chir21}
\end{eqnarray}
Using \eqref{chir20} and \eqref{chir21}, equation \eqref{KillingN53} yields
\begin{eqnarray}
D_{\alpha}\eta^+=2f^{-1}D_{\beta}f\Gamma_{\alpha}^{~~\beta}\eta^+-\frac{1}{2}f^{-1}D_{\alpha}f\eta^+
\label{chir1}
\end{eqnarray}
and
\begin{eqnarray}
D_{\alpha}\eta^-=\frac{1}{2}f^{-1}D_{\alpha}f\eta^-~.
\label{chir2}
\end{eqnarray}
Let us perform two different conformal transformations on the spinors $\eta_+$ and $\eta_-$ as follows
\begin{eqnarray}
\hat{\eta}^+=f^{\frac{1}{2}}\eta^+~,~~~~\hat{\eta}^-=f^{-\frac{1}{2}}\eta^-~.
\label{chir24}
\end{eqnarray}
Implementing \eqref{chir24} into \eqref{chir1} and \eqref{chir2}, we find
\begin{eqnarray}
D_{\alpha}\hat{\eta}^+=2f^{-1}D_{\beta}f\Gamma_{\alpha}^{~~\beta}\hat{\eta}^+
\label{chir3}
\end{eqnarray}
and
\begin{eqnarray}
D_{\alpha}\hat{\eta}^-=0
\label{chir4}
\end{eqnarray}
respectively. Equation \eqref{chir4} implies that
\begin{eqnarray}
\hat{\eta}^-=\sigma^-
\label{chir233}
\end{eqnarray}
where $\sigma^-$ is a 32-component Majorana spinor independent of $s$ which is covariantly constant on $N_4$, i.e.
\begin{eqnarray}
\buildrel{\circ} \over \nabla_ I \sigma^-=0~.
\label{chir25}
\end{eqnarray}
Let us now analyze \eqref{chir3}. The 5-component of \eqref{chir3} yields
\begin{eqnarray}
\frac{\partial \hat{\eta}^+}{\partial s}=\frac{1}{2(c^2s^2+P(z))}\buildrel{\circ} \over\nabla_I P\Gamma^1\Gamma^I  \hat{\eta}^+
\label{chir6}
\end{eqnarray}
where we have implemented \eqref{fszizi2}. We shall explicitly integrate this condition by setting
\begin{eqnarray}
\hat{\eta}^+=\exp\bigg(u(s,z)\buildrel{\circ} \over \nabla_I P\Gamma^1\Gamma^I\bigg)\phi^+~,~~~~~~~~\frac{\partial \phi^+}{\partial s}=0 \ .
\label{chir7}
\end{eqnarray}
On substituting this into \eqref{chir6}, one obtains
\begin{eqnarray}
\frac{\partial u}{\partial s}=\frac{1}{2(c^2 s^2+P(z))}~.
\label{chir10}
\end{eqnarray}
Notice that $P$ is positive by means of \eqref{usefulPP2}. Hence, a solution to equation \eqref{chir10} is given by
\begin{eqnarray}
u(s,z)=\frac{1}{2c\sqrt{P(z)}}\arctan\bigg(\frac{cs}{\sqrt{P(z)}}\bigg)~.
\label{chir11}
\end{eqnarray}
Moreover, equation \eqref{chir7} implies
\begin{eqnarray}
\hat{\eta}^+=\bigg\{\cos\bigg(2c\sqrt{P(z)} u(s,z)\bigg)\mathbb{I}+\frac{1}{2c\sqrt{P(z)}}\sin\bigg(2c\sqrt{P(z)}u(s,z)\bigg)\buildrel{\circ} \over \nabla_I P\Gamma^1\Gamma^I \bigg\}\phi^+~.
\nonumber \\
\label{chir12}
\end{eqnarray}
Inserting \eqref{chir11} into \eqref{chir12}, we get
\begin{eqnarray}
\hat{\eta}^+=\frac{1}{\sqrt{c^2s^2+P(z)}}\bigg(\sqrt{P(z)}\mathbb{I}+\frac{s}{2\sqrt{P(z)}}\buildrel{\circ} \over \nabla_I P\Gamma^1\Gamma^I\bigg)\phi^+~.
\nonumber \\
\label{chir13}
\end{eqnarray}
The $I$-component of \eqref{chir3} is given by
\begin{eqnarray}
\buildrel{\circ} \over \nabla_I \hat{\eta}^+=\frac{c^2s}{c^2s^2+P(z)}\Gamma_I\Gamma^1\hat{\eta}^++\frac{1}{2(c^2s^2+P(z))}\buildrel{\circ} \over \nabla_J P\Gamma_I^{~~J}\hat{\eta}^+~.
\label{chir14}
\end{eqnarray}
Inserting \eqref{chir13} into \eqref{chir14} and using \eqref{usefulhij2}, we get
\begin{eqnarray}
\buildrel{\circ} \over \nabla_I \sigma^+=0
\label{chir16}
\end{eqnarray}
where $\sigma^+$ is given by
\begin{eqnarray}
\phi^+=\frac{1}{\sqrt{P(z)}}\Gamma^I\buildrel{\circ} \over \nabla_I P\sigma^+~.
\label{chir15}
\end{eqnarray}
Inserting \eqref{chir15} in \eqref{chir13}, we obtain
\begin{eqnarray}
\hat{\eta}^+=\frac{1}{\sqrt{c^2s^2+P(z)}}\bigg(\Gamma^I\buildrel{\circ} \over \nabla_I P+2c^2s\Gamma^5\bigg)\sigma^+~.
\label{chir17}
\end{eqnarray}
Moreover, \eqref{KillingN52} is equivalent to
\begin{eqnarray}
\psi=\eta^++\eta^-+\frac{1}{\sqrt{2k}}t(c^2s^2+P(z))^{-\frac{1}{4}}\bigg(2c^2s\Gamma^5+\Gamma^I\buildrel{\circ} \over \nabla_I P\bigg)\Gamma^6\eta^+
\label{chir22}
\end{eqnarray}
where we have used \eqref{chir20} and \eqref{chir21}. Using \eqref{chir24}, \eqref{chir233} and \eqref{chir17}, equation \eqref{chir22} yields
\begin{eqnarray}
\psi&=&2^{-\frac{1}{4}}\big(c^2s^2+P(z)\big)^{-\frac{5}{8}}\bigg(2c^2s\Gamma^5+\Gamma^I\buildrel{\circ} \over \nabla_I P\bigg)\sigma^+ +2^{\frac{1}{4}}\big(c^2s^2+P(z)\big)^{\frac{1}{8}}\sigma^-
\nonumber \\
&-&\frac{2^{\frac{5}{4}}c^2}{\sqrt{k}}t\big(c^2s^2+P(z)\big)^{\frac{1}{8}}\Gamma^6\sigma^+~.
\label{chir40}
\end{eqnarray}
Defining $\check{\sigma}^+:=2^{-\frac{1}{4}}\sigma^+$ and $\check{\sigma}^-:=2^{\frac{1}{4}}\sigma^-$ and dropping the check for simplicity, \eqref{chir40} is equivalent to
\begin{eqnarray}
\psi&=&\big(c^2s^2+P(z)\big)^{-\frac{5}{8}}\bigg(2c^2s\Gamma^5+\Gamma^I\buildrel{\circ} \over \nabla_I P\bigg)\sigma^+ +\big(c^2s^2+P(z)\big)^{\frac{1}{8}}\sigma^-
\nonumber \\
&-&2c^2\sqrt{\frac{2}{k}}~t\big(c^2s^2+P(z)\big)^{\frac{1}{8}}\Gamma^6\sigma^+~.
\label{psic1}
\end{eqnarray}
Let us count the number of supersymmetries preserved by these backgrounds. To this end, define
\begin{eqnarray}
\mathcal{S}^{\pm}:=\textrm{span}\{\sigma^{\pm}\}
\end{eqnarray}
where $\sigma^{\pm}$ satisfy $\Gamma^6\Gamma^{(7)}\sigma^{\pm}=\pm \sigma^{\pm}$ and
\begin{eqnarray}
\buildrel{\circ} \over \nabla_I \sigma^{\pm}=0~.
\end{eqnarray}
Notice that if $\sigma^{\pm}\in\mathcal{S}^{\pm}$, then $\Gamma_{\mu\nu}\sigma^{\pm}\in\mathcal{S}^{\pm}$, where $\mu,\nu$ denote the de-Sitter directions. Hence, using the argument of Section 2.2 of \cite{Farotti:2022xsd}, it follows that
\begin{eqnarray}
\textrm{dim}~\mathcal{S}^{\pm}=8k^{\pm}~,~~~k^{\pm}=1,2~.
\label{count3}
\end{eqnarray}
Moreover, if $\sigma^+\in\mathcal{S}^+$, then $\Gamma_{\mu}\sigma^+\in\mathcal{S}^-$, hence
\begin{eqnarray}
\textrm{dim}~\mathcal{S}^{+}=\textrm{dim}~\mathcal{S}^{-}
\label{observation}
\end{eqnarray}
that is $k^+=k^-:=k$. Using \eqref{psic1}, \eqref{count3} and \eqref{observation}, it follows that the number of supersymmetries is
\begin{eqnarray}
N=\textrm{dim}~\mathcal{S}^{+}+\textrm{dim}~\mathcal{S}^{-}=16k~~~~~~k=1,2~.
\label{count1}
\end{eqnarray}
Notice that equation \eqref{count1} implies that in this class of solutions there are no backgrounds preserving exactly $N=24$ supersymmetries.

To summarize, for this class of solutions, the metric is given by
\begin{eqnarray}
ds^2(M_{11})=H^{-\frac{1}{3}}ds^2(\mathbb{R}^{1,5})+H^{\frac{2}{3}}ds^2(\mathbb{R}\times N_4)
\label{metricH}
\end{eqnarray}
where $N_4$ is a hyperK\"ahler manifold and $H$ satisfies \eqref{KillingN55}, that is
\begin{eqnarray}
-3HD_{\alpha}D_{\beta}H+5D_{\alpha}HD_{\beta}H-(DH)^2\delta_{\alpha\beta}=0~.
\label{hessianH}
\end{eqnarray}
where the functions $f$ and $H$ are related by $f=H^{-{1 \over 6}}$.
In particular, \eqref{hessianH} implies that $H$ is harmonic on $\mathbb{R} \times N_4$, \eqref{harmonicH}. Moreover, the 4-form is given by \eqref{4formH} and is covariantly constant with respect to the Levi-Civita connection of 11-dimensional spacetime. In this case, it follows that the geometry is that of a generalized $M5$-brane configuration, for which the transverse space is ${\mathbb{R}} \times N_4$, and the harmonic function $H$ on $\mathbb{R} \times N_4$ is determined in terms of a hyper-K\"ahler potential $P$ on $N_4$ via 
\begin{eqnarray}
H=\frac{1}{8}\big(c^2s^2+P(z)\big)^{-\frac{3}{2}}
\label{HP}
\end{eqnarray}
where $P$ satisfies \eqref{usefulhij2}.

\subsection{$\mathcal{H}_{\alpha\beta}\ne 0$}

If $\mathcal{H}_{\alpha\beta}\ne 0$, then \eqref{KillingN54} yields
\begin{eqnarray}
(\Gamma^6+\Gamma^{(7)})\eta=0
\label{count10}
\end{eqnarray}
hence $\eta=\eta^-$. Using \eqref{count10}, \eqref{KillingN53} simplifies to
\begin{eqnarray}
D_{\alpha}\chi^-=0~.
\label{aux3491}
\end{eqnarray}
where we have defined $\chi^-:=f^{-\frac{1}{2}}\eta^-$. Equation \eqref{aux3491} implies that
\begin{eqnarray}
\chi^-=\sigma^-
\label{chir23}
\end{eqnarray}
where $\sigma^-$ is a spinor independent of $s$ which satisfies
\begin{eqnarray}
\buildrel{\circ} \over \nabla_ I \sigma^-=0~.
\label{chir25}
\end{eqnarray}
Using \eqref{chir23}, equation \eqref{KillingN52} yields
\begin{eqnarray}
\psi=f^{\frac{1}{2}}\sigma^-~.
\label{chir26}
\end{eqnarray}
Equation \eqref{count3} and \eqref{chir26} imply that the number of supersymmetries preserved by these backgrounds is
\begin{eqnarray}
N=8k^{-}~,~~~~k^{-}=1,2~.
\label{countfinal}
\end{eqnarray}
For this class of solutions, the metric tensor is given by 
\begin{eqnarray}
ds^2(M_{11})=H^{-\frac{1}{3}}ds^2(\mathbb{R}^{1,5})+H^{\frac{2}{3}}ds^2(\mathbb{R}\times N_4)
\label{metricc}
\end{eqnarray}
where $N_4$ is a hyperK\"ahler manifold and $H$ is harmonic on $\mathbb{R}\times N_4$. i.e. it satisfies \eqref{harmonicH}. Moreover, the 4-form is given by \eqref{4formH}. \\
\indent
Notice that equation \eqref{countfinal} implies that $N=24$ is excluded in this class of solutions as well, hence there are no  $D=11$ warped $dS_5$ backgrounds preserving exactly $N=24$ supersymmetries. Moreover, the case $N=8$ has already been analyzed in \cite{Farotti:2022xsd}. Hence, we are left to consider $N=16$. In this case, \eqref{countfinal} implies that there are 16 linearly independent negative chirality spinors $\sigma^-$ which are covariantly constant on $N_4$. Using \eqref{observation}, it follows that there are also 16 positive chirality spinors $\sigma^+$ which are covariantly constant on $N_4$. Hence there are 32 linearly independent covariantly constant spinors on $N_4$. This implies $N_4=\mathbb{R}^4$, and \eqref{metricc}, \eqref{4formH} and \eqref{harmonicH} yield
\begin{eqnarray}
&&ds^2(M_{11})=H^{-\frac{1}{3}}ds^2(\mathbb{R}^{1,5})+H^{\frac{2}{3}}ds^2(\mathbb{R}^5)
\nonumber \\
&&F=\star_5 dH
\nonumber \\
&&\Box_5 H=0~.
\label{M5brane}
\end{eqnarray}
The configuration \eqref{M5brane} corresponds to the standard M5-brane, which indeed preserves 16 supersymmetries in the bulk \cite{Gueven:1992hh}.

\section{Conclusion}

In this work, we have fully classified the warped product $dS_5$ backgrounds in $D=11$ supergravity with enhanced supersymmetry. It is known from \cite{Farotti:2022xsd}
that supersymmetric warped product backgrounds must preserve
$N=8k$ supersymmetries for $k=1,2,3,4$. Our analysis has established the following results for the different possible proportions of supersymmetry:

\begin{itemize}

\item{$N=8$:} The $N=8$ solutions were classified in \cite{Farotti:2022xsd}, and the results for this case are summarized in Section 2. The geometries are generalized $M5$-brane solutions, for which the transverse space is $\mathbb{R} \times N_4$, where $N_4$ is a hyper-K\"ahler manifold.

\item{$N=16$:} There are two possibilities for $N=16$ supersymmetry, corresponding to whether $\mathcal{H}_{\alpha\beta}=0$, where $\mathcal{H}_{\alpha\beta}$ is
defined in ({\ref{definHIJ}}). 

\begin{itemize}

\item[(i)] If $\mathcal{H}_{\alpha\beta}=0$ then the $N=16$ solutions have the property that the
4-form $F$ is covariantly constant with respect to the 11-dimensional Levi-Civita connection and is given by \eqref{4formH}. Such solutions have been discussed in Section 5.4 of \cite{Farotti:2022xsd}; here we establish that these solutions actually have enhanced supersymmetry.
 The geometry corresponds to a generalized M5-brane configuration \eqref{metricH}, with transverse space ${\mathbb{R}} \times N_4$ for which the harmonic function $H$ on $\mathbb{R} \times N_4$ is determined in terms of a hyper-K\"ahler potential $P$ on $N_4$ via \eqref{HP}. 

\item[(ii)] If however, $\mathcal{H}_{\alpha\beta}\ne0$ then we find that the bosonic fields are given by \eqref{M5brane}. This configuration corresponds to a stack of M5-branes with transverse space $\mathbb{R}^5$ \cite{Gueven:1992hh}.

\end{itemize}

\item{$N=24$:} There are no warped product $dS_5$ solutions preserving exactly $N=24$ supersymmetries.

\item{$N=32$:} This case has been considered in \cite{Farotti:2022xsd} and it has been shown that the only possibilities are $\mathbb{R}^{1,10}$ with vanishing 4-form, or the maximally supersymmetric $AdS_7 \times S^4$ solution.

\end{itemize}

\setcounter{section}{0}
\setcounter{subsection}{0}

\section*{Acknowledgments}

DF is partially supported by the STFC DTP Grant ST/S505742.

\section*{Data Management}

No additional research data beyond the data presented and cited in this work are needed to validate the research findings in this work.

\end{document}